\documentclass{PoS}

\title{Continuing EVN monitoring of HST-1 in the jet of M87}

\ShortTitle{EVN monitoring of HST-1}

\author{\speaker{Kazuhiro Hada}$^{1,2}$, Marcello Giroletti$^{1}$, Gabriele
                  Giovannini$^{1,3}$, Carolina Casadio$^{4}$, Matthias
                  Beilicke$^{5}$, Andrea Cesarini$^{6}$, Teddy Cheung$^{7}$,
                  Akihiro Doi$^{8}$, Jos\'e Luis G\'{o}mez$^{4}$, Henric Krawczynski$^{5}$, Motoki Kino$^{9}$ and Hiroshi Nagai$^{2}$\\ }

\author{\\
        $^1$INAF Istituto di Radioastronomia, Via Gobetti 101, 
	    40129 Bologna, Italy\\
	$^2$National Astronomical Observatory of Japan, 
	    2-21-1 Osawa, Mitaka, 181-8588 Tokyo, Japan\\
	$^3$Dipartimento di Fisica e Astronomia, Universit\'{a} di Bologna, via Ranzani 1, I-40127 Bologna, Italy
	$^4$Instituto de Astrofisica de Andalucia, CSIC, 
	    Apartado 3004, 18080 Granada, Spain\\
	$^5$Washington University, St. Louis, MO 63130, USA\\
	$^6$Department of Physics, University of Trento, I-I38050 Povo, Trento, Italy\\
	$^7$Space Science Division, Naval Research Laboratory, Washington, 
	    DC 20375-5352, USA \\
	$^8$Institute of Space and Astronautical Science, 3-1-1 Yoshinodai, 
	    Sagamihara, 229-8510 Kanagawa, Japan\\
	$^9$Korea Astronomy and Space Science Institute (KASI), 776 Daedeokdae-ro, Yuseong-gu, Daejeon 305-348, Korea\\
	    E-mail: \email{kazuhiro.hada@nao.ac.jp}
	}

\abstract{The relativistic jet in M87 offers a unique opportunity for understanding the detailed jet structure and emission processes due to its proximity. In particular, the peculiar jet region HST-1 at $\sim$1 arcsecond (or 80\,pc, projected) from the nucleus has attracted a great deal of interest in the last decade because of its superluminal motion and broadband radio-to-X-ray outbursts, which may be further connected to the $\gamma$-ray productions up to TeV energies. 
Over the last five years, we have been doing an intensive monitoring of HST-1 with EVN at 5\,GHz in order to examine the detailed structural evolution and its possible connection to high-energy activities. While this program already yielded interesting results in terms of the detailed mas-scale structure, proper motion measurements and structural variations, the recent HST-1 brightness is continuously decreasing at this frequency. To counter this, we have shifted our monitoring frequency to 1.7\,GHz from October 2013. This strategy successfully recovered the fainter emission that was missed in the last 5\,GHz session. Moreover, we again discovered the sudden emergence of a new component at the upstream edge of HST-1, demonstrating that the use of EVN 1.7GHz is indeed powerful to probe the current weak nature of HST-1. Here we report early results from the 1.7\,GHz monitoring as well as further progress on the long-term kinematic study.}

\FullConference{12th European VLBI Network Symposium and Users Meeting,\\
                 7-10 October 2014\\
                 Cagliari, Italy}

\begin{document}

\section{Introduction}
Understanding of the formation of Active-Galactic-Nuclei (AGN) jets 
as well as their connection to
$\gamma$-ray productions up to tera-electronvolt (TeV) energy is one of the
primary goals in current high-energy astrophysics. The nearby radio galaxy M87 is
one of the best studied AGN jets through radio to TeV $\gamma$-ray, and especially
VLBI observations are able to access the jet structure on its formation and
collimation scale thanks to the proximity ($1~{\rm mas}=0.08~{\rm pc}=140~R_{\rm
s}$ for $D=16.7$ Mpc and $M_\mathrm{BH}= 6 \times 10^9 M_\odot$), allowing us to
obtain detailed observational constraints on relativistic-jet phenomena at an
unprecedented linear resolution [1, 2, 3, 4, 5, 6].

At a distance of $\sim$120 pc or $\sim$$4.8\times10^{5}~R_{\rm s}$ from the
nucleus, the M87 jet exhibits a remarkable feature known as HST-1~[7]. Probing this
feature provides a clue to the above key question on AGN jets. In 2005, a large
TeV flare from M87 was accompanied by the radio-to-X-ray outbursts from HST-1,
subsequently with the emergence of superluminal features~[8]. 
On the other hand, HST-1 seems to be located in a `transition zone' of the M87's
jet-collimation profile, where the jet shape changes from a parabolic (at
$r\lesssim 10^{5}R_{\rm s}$) to a conical one (at $r\gtrsim 10^{6}R_{\rm s}$),
together with a possible smaller cross section on HST-1 itself~[9]. 
These results leads to an
interesting hypothesis that HST-1 marks a recollimation 
shocked region at the end of the acceleration and collimation zone, resulting in a
significant energy release in the form of TeV flares [8, 9, 10]. However, the exact kinematics of HST-1 and its mas-scale structure remain uncertain. To better understand these properties and their possible connection to the $\gamma$-ray production, it is necessary to keep a long-term, continuous VLBI monitoring. 

In this context, from mid 2009 we started a detailed monitoring program of HST-1
with EVN at 5\,GHz~[11]. As reported in Giroletti et al.~[12], this project
already provided interesting fidings on the HST-1 kinematics by combining the
earlier (dating back to 2006) VLBA archival data at 1.7\,GHz; (i) we revealed a
detailed mas-scale structure and evolution, where the HST-1 complex was resolved
into several compact subcomponents; (ii) we determined the individual proper
motions for these components, where most of them are moving superluminally over 5
years; (iii) moreover in 2010, we discovered the ejection of a new feature from
the upstream edge of HST-1, which occurred coincidentally with the large TeV flare
from M87. These results imply that HST-1 can be actually associated with the
high-energy $\gamma$-ray production.

\begin{figure}[ttt]
\begin{center}
 \scalebox{0.5}{\includegraphics{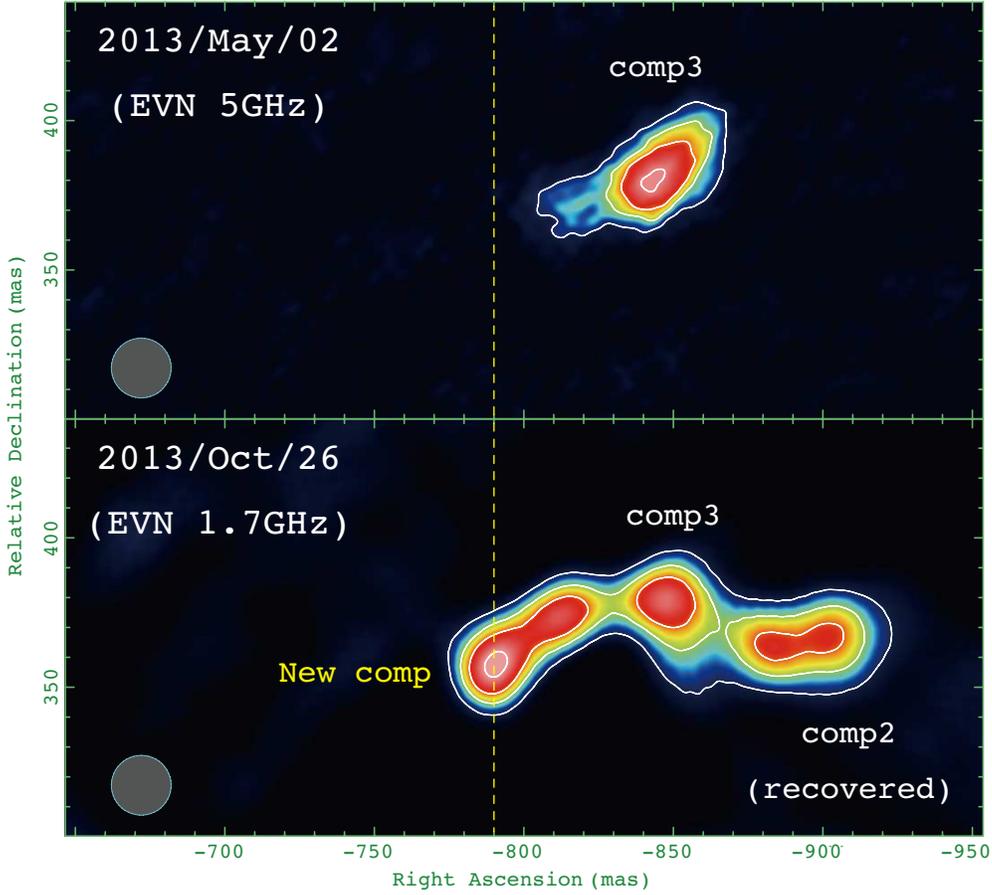}} 
 \caption{Comparison of two recent EVN images. The upper image is obtained at the last epoch of the 5-GHz campaign (May 2013), while the lower panel shows a 1.7~GHz image taken at the first epoch of the 1.7\,GHz monitor (October 2013).} \label{fig:}
\end{center}
\end{figure}

\section{EVN monitoring of HST-1 at 1.7GHz; preliminary results}
While the EVN 5\,GHz program has been successful, the recent HST-1 brightness is continuously decreasing. To counter this and further continue our monitoring, we have shifted our monitoring frequency to 1.7~GHz from the recent session in October 2013 because HST-1 shows a steep radio spectrum~[13]. Fig.\,1 shows a comparison of recent two HST-1 images. The top panel indicates an image obtained at the last epoch of the 5\,GHz monitor while the bottom one was taken at the first epoch of the 1.7\,GHz program.
We successfully recovered the fainter emission that was missed in the last 5~GHz session. Moreover, we again discovered the sudden emergence of a new component at the upstream edge of HST-1 (here we call it comp4). These results demonstrate that the use of EVN 1.7~GHz is indeed powerful to investigate the current weak nature of HST-1. 

In Fig.\,2 we show an updated distance-versus-time plot of HST-1 subcomponents from the core. The long-living components (i.e., comp1, comp2 and comp3) are still continuously moving superluminally. A constant-speed fitting over the entire period results in 4.5\,$c$ for comp1 and comp2, while comp3 has a slightly faster speed of 5.1\,$c$.  
The new component (comp4) is also moving at a superluminal speed, and a tentative estimate of the apparent speed with the 3 data points results in $\sim$10\,$c$, although the uncertainty is still very large ($\pm\sim$5\,$c$).  

Looking into the kinematics of the long-living components in more detail, we note that their motions are not following simple straight lines, which means that the actual kinematics are more complicated. In Fig.\,3 we show the observed two-dimensional trajectories on the sky for these three components. As evident in this figure, while all the components are ejected from a similar location, their subsequent trajectories are remarkably different from one another. All the trajectories are never ballistic but smoothly curved, and then we detected significant proper motions not only along but also in the direction perpendicular to the overall jet axis (i.e., PA$\sim$$203^{\circ}$).  
For comp1 and comp2, the observed position displacements in this direction since their emergences are as large as 30\,mas, 
indicating an apparent speed of $\sim$1\,$c$. 
On the other hand, comp3 is traveling differently; it was initially ejected towards PA$\sim$$310^{\circ}$, and subsequently showed a remarkable turnaround, changing the  direction towards the jet axis. The onset of a turnaround is also seen at the latest stage of comp2. Moreover, it seems that these components are oscillating also in their apparent speeds on a time scale of several years. This oscillation tends to occur synchronously with the change of the moving direction.

\begin{figure}[ttt]
\begin{center}
 \scalebox{0.8}{\includegraphics{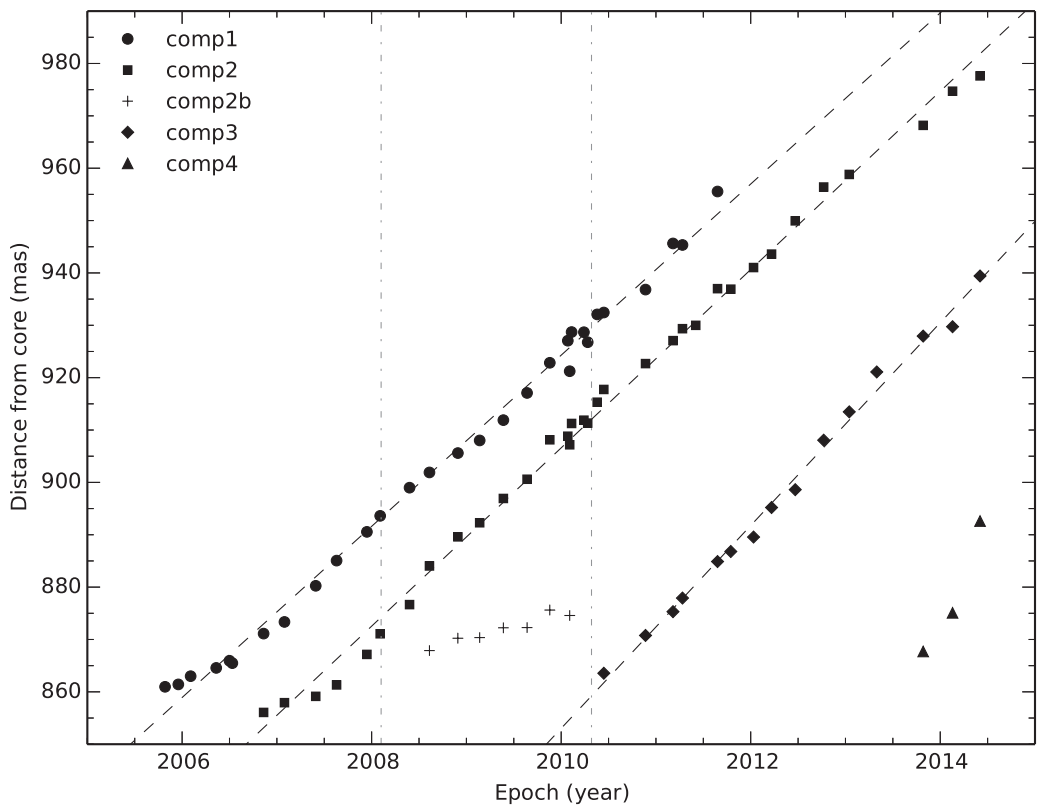}} 
 \caption{Updated distance of the HST-1 subcomponents from the core versus time. The observations at the last three epochs are made at 1.7\,GHz with EVN. The dashed lines show least-squares linear fits for comp1, comp2 and comp3, while the vertical dot-dashed lines represent the epochs of the 2008 and 2010 VHE flares.} \label{fig:}
\end{center}
\end{figure}

\section{Summary and conclusion}

Nearly a decade-long VLBI monitoring of HST-1 is beginning to reveal the detailed time evolution of the kinematics for the resolved substructures, where we found some changes (and slow oscillations) in both direction and speed with time (and their possible correlation). This suggests that the actual trajectories of these components are three-dimentional. If this is the case, a likely scenario is that the HST-1 complex is traveling along a helical path at a relativistic speed. Interestingly, a recent VLA polarimetric study of HST-1 independently detected a progressive rotation of HST-1's EVPA during a part of our monitoring period~[14], which could be related the observed curved trajectories. To further test this scenario, it would be fruitful to make a EVPA monitor for the resolved structure with VLBI or to search for a possible year-scale periodicity in the motion by further continuing our monitoring. 

We finally note that the HST-1 monitor with EVN is also useful for understanding the overall formation and collimation mechanisms of M87 jet. It is suggested that the HST-1 cross section is smaller than that expected from the parabola-shape collimation profile of the inner jet [9]. However as shown in Fig.\,1, the recent HST-1 becomes significantly larger ($\sim$double) in size - not only along but also across the jet - than that seen in the past. 
This indicates that the underlying HST-1 structure should be much more extended than previously thought, and the individual bright components may occupy only a small part of the entire jet cross section. In order to examine this issue, we have recently conducted a deep EVN 1.7\,GHz session incorporating the eMERLIN, which greatly improves the detectability for the surrounding, more extended structure. This will allow us to uncover the entire jet structure around HST-1 in more detail.

\begin{figure}[ttt]
\begin{center}
 \scalebox{0.8}{\includegraphics{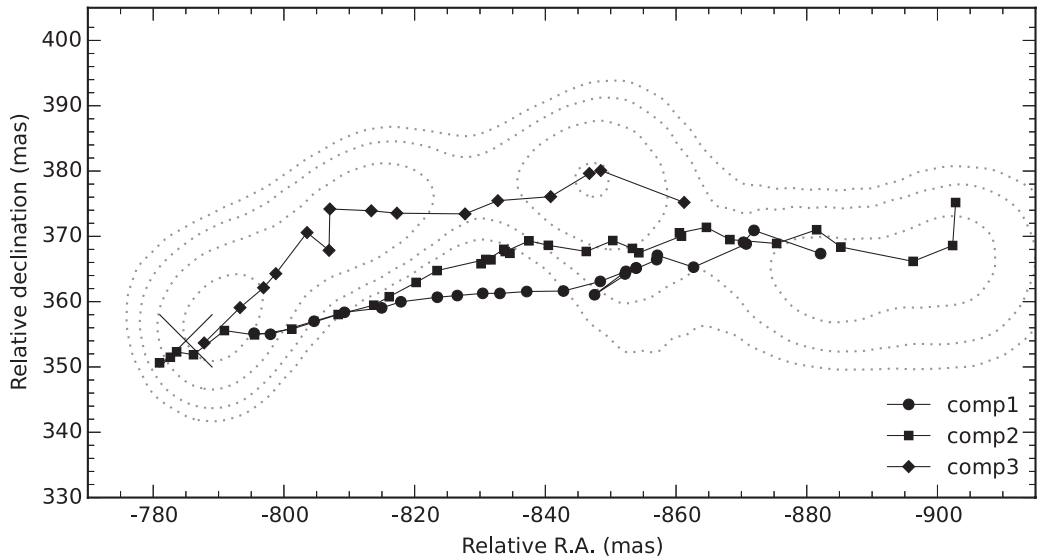}} 
 \caption{Observed (x,y) trajectories for comp1, comp2 and comp3. The underlying contour with dotted lines is a EVN 1.7\,GHz image obtained in October 2013. } \label{fig:}
\end{center}
\end{figure}

\bigskip
\noindent
\textbf{Acknowledgments.} We acknowledge a contribution from the Italian Foreign
Affair Minister under the bilateral scientific collaboration between Italy and
Japan. We acknowledge financial contribution from grant PRIN-INAF-2011. e-VLBI research infrastructure in Europe is supported by the European
Union's Seventh Framework Programme (FP7/2007-2013) under grant agreement
no. RI-261525 NEXPReS.  The European VLBI Network is a joint facility of European,
Chinese, South African and other radio astronomy institutes funded by their
national research councils. The National Radio Astronomy Observatory is a facility
of the National Science Foundation operated under cooperative agreement by AUI.

\end{document}